\documentclass{osa-article}

\journal{oe}

\articletype{Research Article}

\begin{document}

\title{Optical detection of nano-particle characteristics using coupling to a nano-waveguide}

\author{Masakazu Sugawara,\authormark{1,*} Yasuyoshi Mitsumori,\authormark{1} Keiichi Edamatsu,\authormark{1} and Mark Sadgrove\authormark{1, 2}}

\address{\authormark{1}RIEC, Tohoku University, Sendai, 980-8577, Japan\\
\authormark{2}Tokyo University of Science, Tokyo, 162-8601, Japan\\}

\email{\authormark{*}sugawara@quantum.riec.tohoku.ac.jp} 



\begin{abstract}
Recently, much research concerning the combination of nano-scale waveguides with nano-crystals and other nano-particles has been reported, because of possible applications in the field of quantum information and communication. The most useful and convenient method to verify the nature of such systems is optical detection. However, due to the diffraction limit, optical identification of characteristics such as particle type, particle position, etc is difficult or impossible. However, if such particles are placed on a waveguide, the coupling of scattered light to the waveguide guided modes can reveal the information about the particles. Here we consider how illumination with light of arbitrary polarization can reveal the difference between isotropic and non-isotropic nano-particles placed on the surface of an optical nanofiber. Specifically, we measure the polarization response function of gold nano-rods (GNRs) on an optical nanofiber surface and show that it is qualitatively different to that for gold nano-spheres (GNSs). This experimental technique provides a simple new tool for the optical characterization of hybrid nano-optical devices.
\end{abstract}

\section{Introduction}
The ability to combine nano-structures to create nano-optical devices with unique characteristics is a vital technique in modern quantum optical research \cite{Benson:2011Nature}. In particular, the low-cost and flexible nature of combining commercial nano-crystals and nano-particles with waveguide technology (in comparison to conventional nanofabrication) has led to a boom in nano-photonics research \cite{Benson:2009NL, barth:2010NanoLett, schell:2017ACSPhotonics, Nayak:2018IOP, Groreux:2018PRAp}. Many of the techniques developed in these references are non-deterministic in the placement of particles or have both deterministic and non-deterministic aspects. In the case where techniques are non-deterministic, it is desirable to characterize the combined structure in-situ optically where possible.

As regards such characterization, the optical detection of sub-wavelength sized nano-particles presents a particular difficulty due to the restriction of the diffraction limit. In particular, for visible or near-visible light, particles in the < ~100 nm size range may have optical responses indistinguishable from point dipoles \cite{BohrenHuffman:2008}.  For the case of photoluminescent nano-crystals such as quantum dots and defect centers in nano-diamond, extra information, e.g. regarding nano-particle number, can be gleaned from photon statistics \cite{englund:2010NanoLett, fujiwara:2011NanoLett, yalla:2012PRL, yalla:2012OptExp, hausmann:2013NanoLett, yalla:2014PRL, liebermeister:2014APL, Schell:2015SciRep}. However, such information is lacking for simple nano-particles. 

For particles placed on a substrate (such as a nano-waveguide) which has its own mode structure, the coupling of scattered light with these modes may allow information about the nano-particle to be retrieved. Important recent examples include the case of nano-particle scatterers coupled to nano-waveguides \cite{Petersen:2014Science, Mark:2017SciRep} in which the homogeneous polarizability of spherical nano-particles allowed the chiral mode structure of nano-waveguides to be revealed using polarization controlled probe light.

Here we consider how illumination with light of arbitrary polarization can reveal the difference between isotropic and non-isotropic nano-particles placed on the surface of a nanofiber.
Specifically, we measure the polarization response function of gold nano-rods on a nanofiber surface and show that it is qualitatively different to that we measured previously\cite{Mark:2017SciRep} for gold nano-spheres. This fundamental difference in the polarization response allows even strongly sub-wavelength sized nano-particles to be optically distinguished based on the characteristics of coupling to the nano-waveguide, and could also allow us to distinguish single particles from particle clusters in principle.

\section*{Results}
\begin{figure}[ht]
	\centering
	\includegraphics[width=\linewidth]{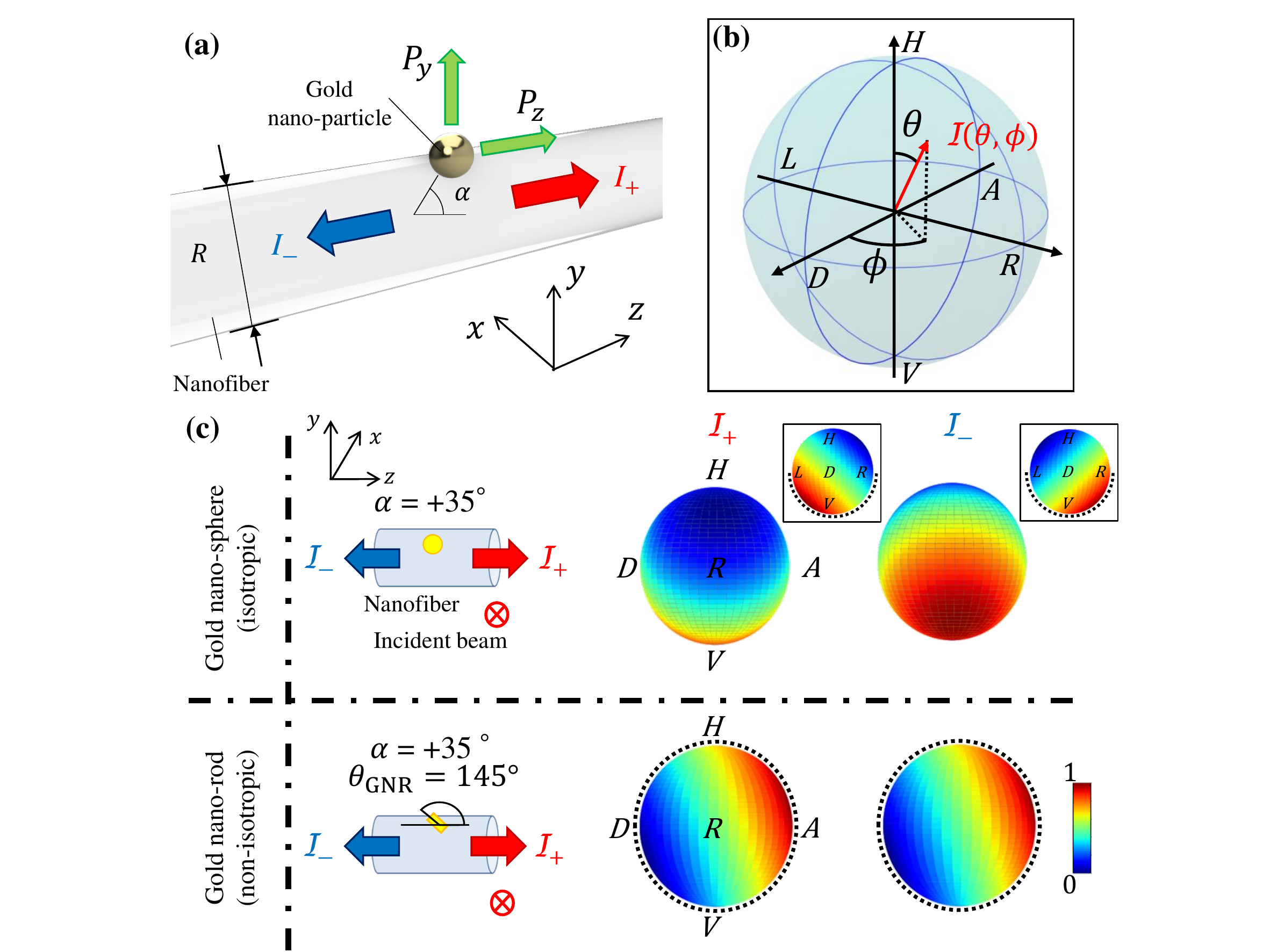}
	\caption{\textbf{(a)} Illustration of a gold nano-particle on an optical nanofiber with a diameter of $R$. The gold nano-particle is placed at the azimuthal angle $\alpha$ [deg.] on the surface of the fiber. $P_{z}$ and $P_{y}$ are complex components of the excitation field after passing through a quarter wave-plate (QWP) and a half wave-plate (HWP). The intensity coupled into the guided modes of the fiber in the $\pm z$ direction is given by $I_{\pm}$. \textbf{(b)} Definition of $\theta$ and $\phi$ on the Poincar\'{e} sphere. The correspondence between the angles of the two wave-plates, $(\beta,\, \gamma)$, and $(\theta,\, \phi)$ is explained in the main text. \textbf{(c)} Examples of the theoretical polarization response function (PRF) of a gold nano-sphere (GNS) and a gold nano-rod (GNR) on the surface of an optical nanofiber. The leftmost figures in each row depict the nano-particle configuration which leads to the displayed PRF. The insets in the upper row are the PRF viewed from $(\theta,\ \phi) =$ (90\textdegree, 0\textdegree), $i.e.$ $D$-polarization. In all cases, calculations were performed for light incident from the $-x$ direction and propagating along the $x$ axis. Note that $\theta_{\rm GNR}$ gives the angle of the nano-rod relative to the horizontal axis. Black dotted lines indicate the positions on the surface of the Poincar\'{e} sphere where the maximum of the PRF can lie in principle.}
	\label{fig:concept}
\end{figure}

\begin{figure}[ht]
	\centering
	\includegraphics[width=\linewidth]{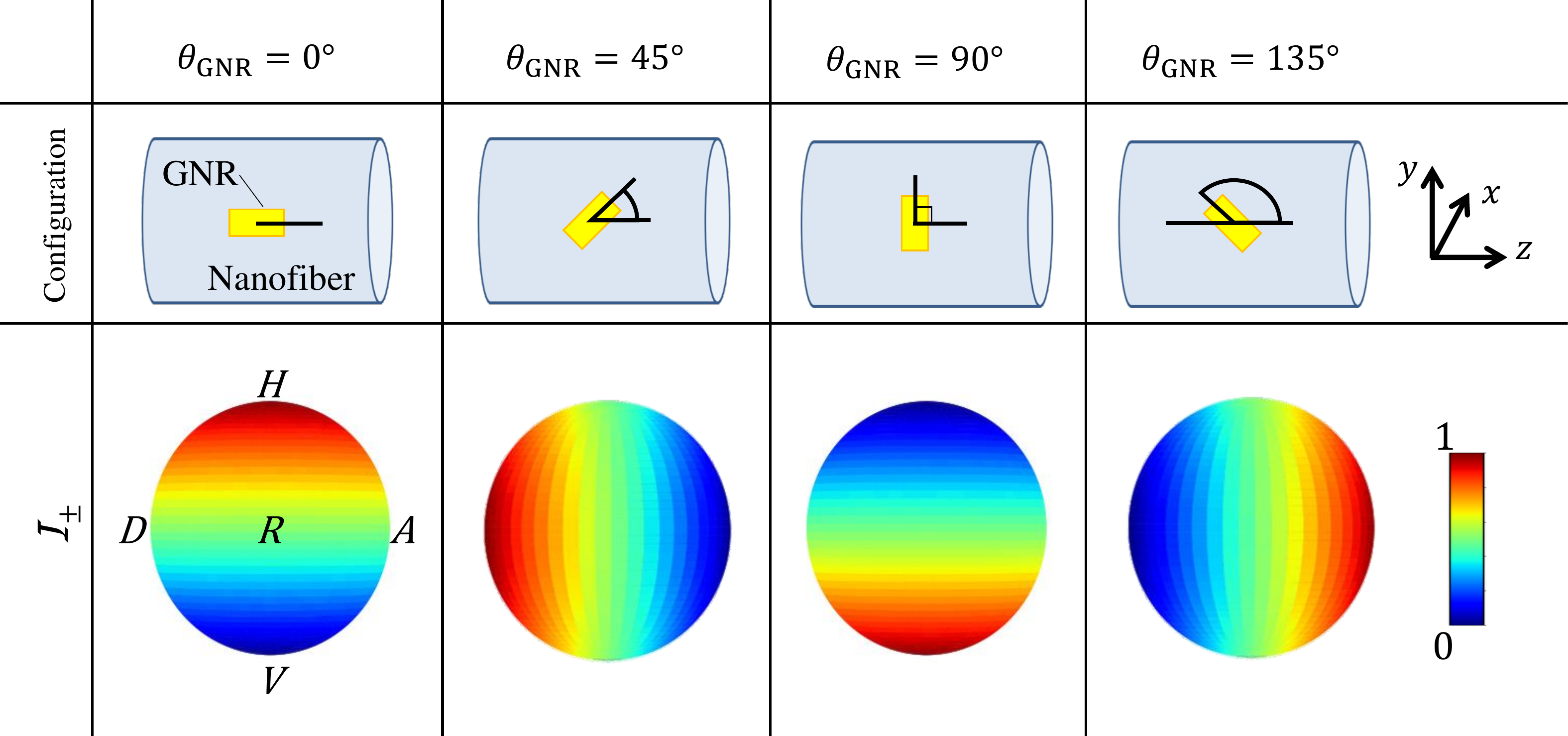}
	\caption{Theoretical calculations of the PRF for gold nano-rods (GNR) at different values of $\theta_{\rm GNR}$ as indicated in the figure. The calculations were performed for a nanofiber radius of 400 nm. The GNRs are situated at azimuthal angle $\alpha$ of 0\textdegree.}
	\label{fig:PRFofGNRTheory}
\end{figure}

We first introduce the theoretical principle of the polarization response function (PRF) for a nano-spheres and nano-rods on the surface of an optical nanofiber. The optical coupling of a dipole scatterer to the guided mode of the optical nanofiber is determined by Hermitian product of the electric dipole moment $\textbf{d}$ of the scatterer on the nanofiber surface and the fundamental guided modes of the nanofiber $\epsilon_{\pm}$ as \cite{Petersen:2014Science},
\begin{equation}\label{eq:Hermitian product}
I_{\pm} \propto \lvert {\bf d} \cdot \epsilon_{\pm}^{*} \rvert^{2},
\end{equation}
where $I_{\pm}$ are the intensities coupled into the guided modes that propagate in opposite ($\pm$) directions along the nanofiber. We note that despite the fact that the nanofiber section of the fiber taper may have a radius larger than that required for the single mode condition, because our fiber tapers are made from single mode fiber, we only measure light coupled to the nanofiber fundamental modes. We also note that in the case of isotropic scatterers it has been found to be sufficient to consider only the $y-$ polarized modes when making the calculation given in Eq. \ref{eq:Hermitian product}. However, in the case of nanorods, we also include coupling to the $x-$ polarized modes in our calculation. In this experiment, the dipole moment of the particle is induced by an external electric field (See Fig. \ref{fig:concept}(a)) and can be greatly enhanced by plasmon-polariton resonance at the particle surface. 
Nano-particles on the nanofiber surface are illuminated by an external field whose polarization state can be varied over the entire surface of the Poincar\'{e} sphere by using a quarter wave plate (QWP) and a half wave plate (HWP) placed in the optical beam path. For the experimental reconstruction of the PRF, $\mathcal{I}$, we first measured the intensities $I_{\pm}(\beta, \gamma)$ detected at the two nanofiber output ports, which is a function of the angles $\beta$ and $\gamma$ of the HWP and QWP, respectively. We initialize the polarization of the incident light to horizontal ($H$) represented as $\mathbf{P}_{\rm in}=[1,0]^T$ in the Jones vector calculus. Then, we can express the polarization state of light after the two wave plates as,
\begin{equation}\label{eq:P_Jones}
\boldsymbol{\mathrm{P}} = \hat{\mathrm{M}}_{\mathrm{HWP}}(\gamma)\hat{\mathrm{M}}_{\mathrm{QWP}}(\beta)\boldsymbol{\mathrm{P}}_{\mathrm{in}},
\end{equation}
where $\hat{\mathrm{M}}_{\mathrm{HWP}}(\gamma)$ and $\hat{\mathrm{M}}_{\mathrm{QWP}}(\beta)$ are the Jones matrices for a half wave plate and a quarter wave plate with the fast axis angle of $\gamma$ and $\beta$. Finally, we project $I_{\pm}(\beta, \gamma)$ to $\mathcal{I}_{\pm}(\theta, \phi)$ on the Poincar\'{e} sphere as shown in Fig. \ref{fig:concept}\textbf{(b)} using the following relations,
\begin{equation}
\theta = 2\mathrm{tan}^{-1}(\lvert P_{z}/P_{y}\rvert),\;\; \phi = \mathrm{arg}(P_{z}) - \mathrm{arg}(P_{y}),\;\; \mathcal{I}_{\pm}(\theta,\, \phi) = I_{\pm}(\beta,\, \gamma),
\end{equation}
where $P_{z}$ and $P_{y}$ are the complex horizontal and vertical components of the polarization state $\boldsymbol{\mathrm{P}}$ and arg(\textit{c}) denotes the angle of the complex number c in the complex plane. In the case of a GNS, we approximate the particle as a point dipole with the same polarization as the input light. In the case of a GNR, to a good approximation, only the plasmonic mode lying along the nano-rod principle axis is excited \cite{ming:2009strong}, meaning that the GNR dipole moment also lies along this direction. In particular, we note that the polarization of the dipole moment of the GNR is independent of the polarization of the incident beam. The strength of the dipole moment ranges from zero if the polarization is perpendicular to the rod axis and maximal if the polarization is aligned with the rod axis. Considering these properties, the PRFs of a GNS and a GNR can be calculated from Eq. \ref{eq:Hermitian product} as shown in Fig. \ref{fig:concept}(c). In this example, both particles are situated at an azimuthal angle $\alpha = $+30\textdegree \ on the surface of the optical nanofiber which has a diameter of 400 nm. The angle of the long axis of the GNR, $\theta_{\rm GNR}$, is set to be 145\textdegree. The incident beam is assumed to propagate along the $x$ axis in the positive direction. The wavelength of the beam is 785 nm. The left and right hand PRFs of the GNS, as shown in the top row of Fig. \ref{fig:concept}(c), are reflections of each other, due to the chiral nature of the coupling from an isotropic scatterer \cite{Petersen:2014Science, Mark:2017SciRep}. On the other hand, the left and right PRFs for a nano-rod are exactly the same due to the fact that the nano-rod only supports a linear dipole moment. Furthermore, the maximum value of the PRF for a GNS should lie on the half great circle of the Poincar\'{e} sphere passing through $L-V-R$ as depicted in the insets of Fig. \ref{fig:concept}(c) by a black dotted line. In particular, we emphasize that the maximum coupling will never occur at the $H$ polarization for a GNS because the $z-$ ($H$ polarized) component of the nanofiber mode is always less than the $y-$ ($V$) polarized component. Contrary to the case of a GNS, the maximum or minimum value of the PRFs of a GNR is theoretically found on the great circle of the Poincar\'{e} sphere passing through $H-D-V-A$. This is because, as mentioned earlier, a single GNR behaves like a polarizer and maximum (minimum) coupling occurs when the incident light has linear polarization along (perpendicular to) the axis of the GNR on the optical nanofiber. We show the examples of the PRFs of single GNRs at different values of $\theta_{\rm GNR}$ in Fig. \ref{fig:PRFofGNRTheory}. It can be seen that maximum (minimum) values are rotated on $H-D-V-A$ great circle with respect to $\theta_{\rm GNR}$.

\begin{figure}[ht]
	\centering
	\includegraphics[width=\linewidth]{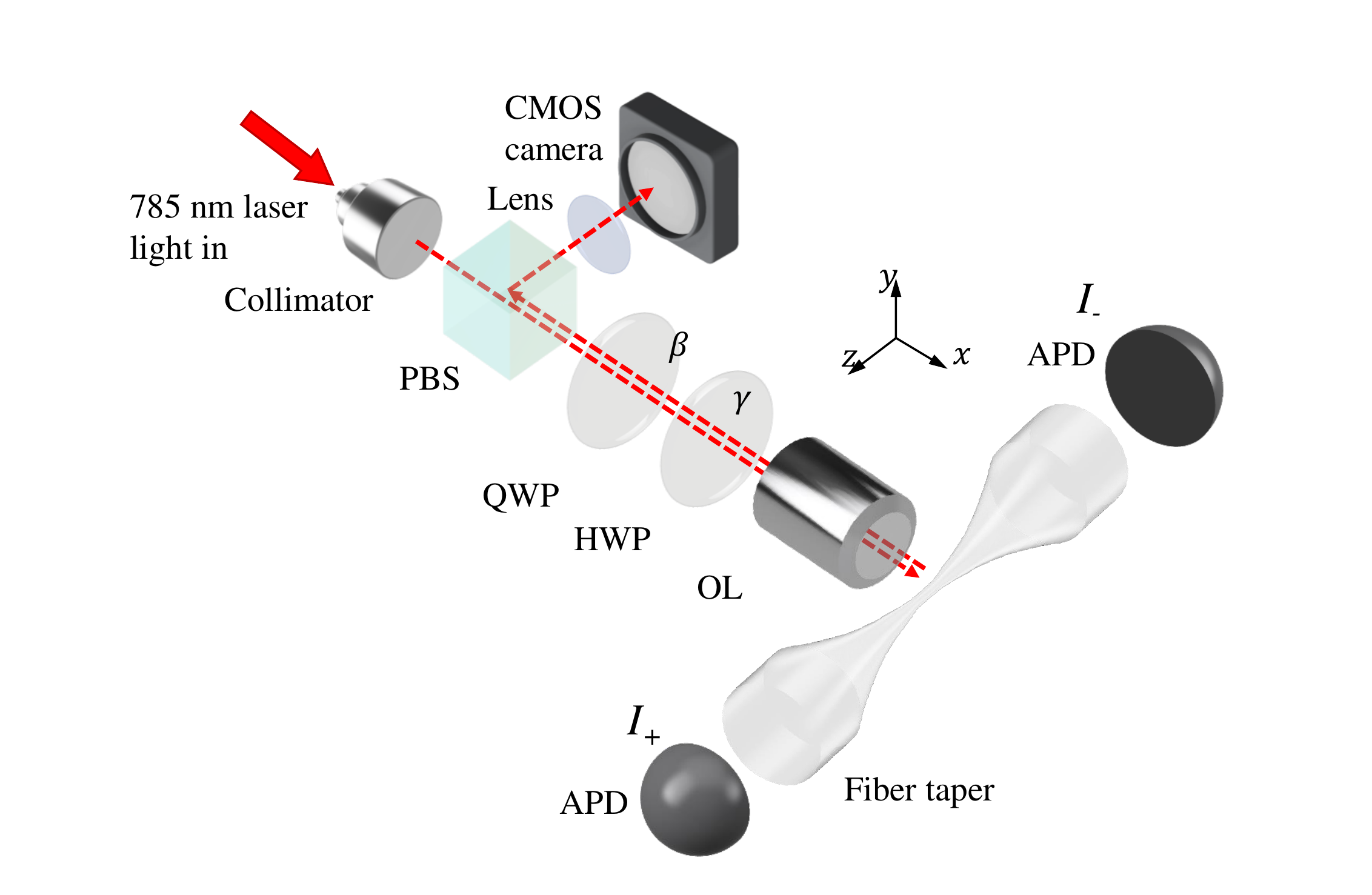}
	\caption{\textbf{(a)} Experimental setup. Light from a free running diode laser ($\lambda$ = 785 nm) is sent via optical fiber to a collimator and passes through a polarizing beam splitter (PBS), quarter wave plate (QWP), and a half wave plate (HWP) as depicted by the broken red lines. The light is focused on the nanofiber surface by an objective lens (OL, 10x magnification). Light coupled into the guided mode of the optical nanofiber is detected by avalanche photodiodes (APD). The complementary metal-oxide semiconductor (CMOS) camera is used to monitor the scattered light from the nanofiber surface and nano-particles on the fiber directly.}
	\label{fig:Setup}
\end{figure}

\begin{figure}[ht]
	\centering
	\includegraphics[width=\linewidth]{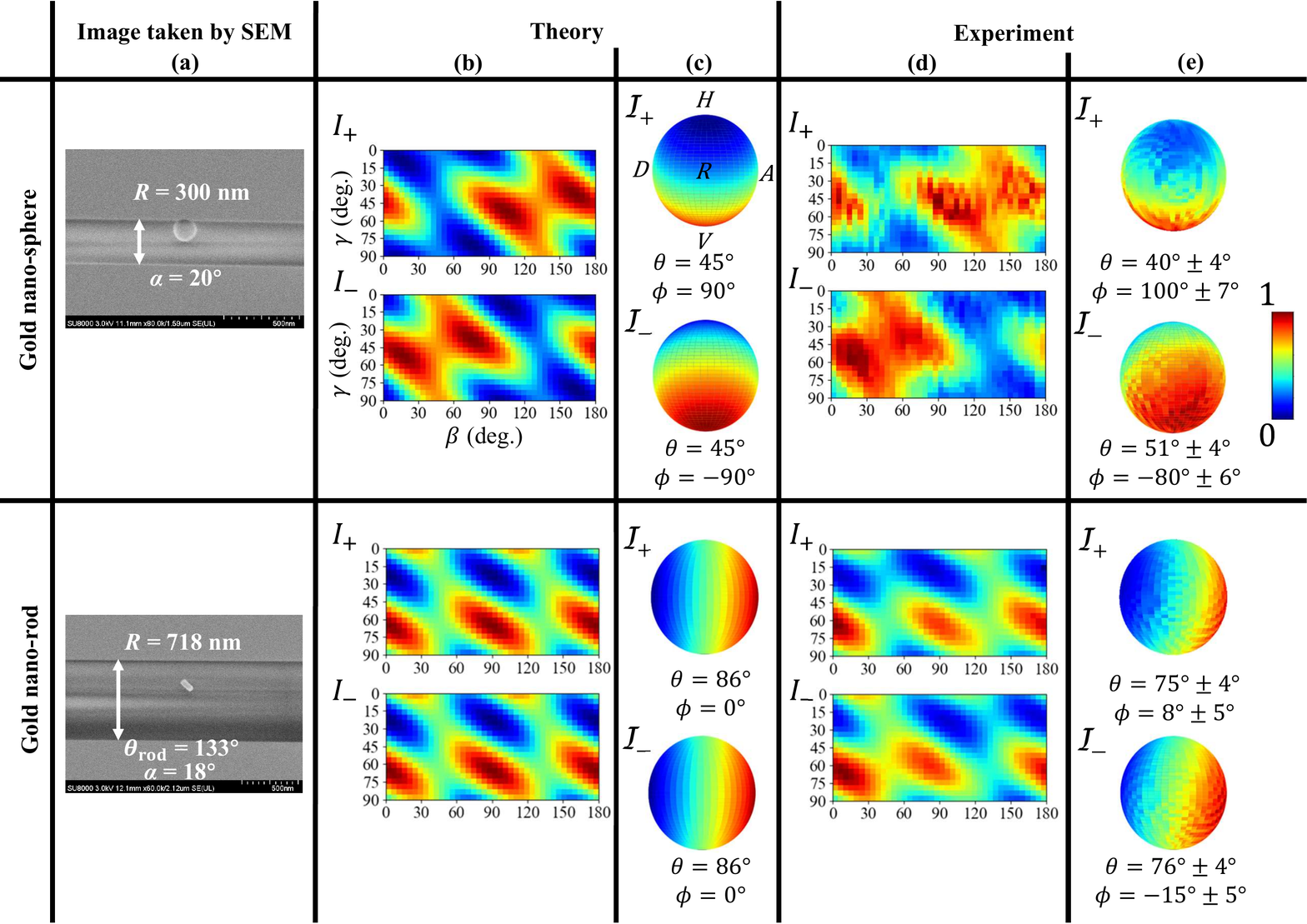}
	\caption{Theoretical calculations and experimental measurements of the PRFs for a single GNS (top row) and a single GNR (bottom row). (\textbf{a}) Images taken by scanning electron microscope (SEM). $R, \alpha, \theta_{\rm GNR}$ are the fiber diameter, the azimuthal angle of the particle and the angle of the GNR, respectively. (\textbf{b}) Intensities $I_{\pm}$ as a function of the angles of  the two wave plates as predicted by theory. (\textbf{c}) PRFs reconstructed from (\textbf{a}). (\textbf{d}) Experimentally obtained data as a function of the angles of the two wave plates. (\textbf{e}) PRFs reconstructed from (\textbf{d}). Angles $\theta$ and $\phi$ indicate positions of minimum value in PRFs.}
	\label{fig:TheoryAndExp}
\end{figure}

Moving on to the experimental measurement of the phenomena discussed above, we used the experimental setup shown in Fig. \ref{fig:Setup} to illuminate gold nano-particles on the surface of an optical nanofiber \cite{Mark:2017SciRep}. We prepared an optical nanofiber with a waist diameter of 400 nm on average. We formed the fiber taper from a commercial single-mode optical fiber which was single mode for wavelenths above 780 nm, creating an effective spatial mode filter and ensuring that only coupling to the fundamental mode was measured \cite{yalla:2012OptExp, yalla:2012PRL}. After that, we deposited GNSs (diameter 150 nm) or GNRs (length 160 nm, diameter 50 nm) using the methods described below. GNSs were deposited using the same method we reported recently \cite{Mark:2017SciRep}. Specifically, we first prepared a droplet of GNS solution near the nanofiber using a micro pipette. We then moved the tungsten needle towards the nanofiber through the droplet until the needle touched the surface of the fiber. We typically achieved deposition of single GNSs within 10 trials using this method. A scanning electron microscope (SEM) image of such a deposition is shown in the top row of Fig. \ref{fig:TheoryAndExp} (\textbf{a}). We found that this method was less successful for GNRs, and we therefore deposited GNRs using a simpler method: We brushed the surface of a nanofiber with a droplet of colloidal GNR solution until scattering was observed at the nanofiber surface. An example of a nano-rod deposition produced by this method is shown in the bottom row of Fig. \ref{fig:TheoryAndExp} (\textbf{a}). After deposition, we perform optical measurements to detect particle position by focusing a laser (wavelength 785 nm) at the surface of the optical nanofiber, and scanning the fiber through the laser spot. When the laser spot overlapped with a nano-particle, a sharp rise in scattering of light into the nanofiber guided modes was detected, using an avalanche photodiode (APD) connected to one end of the fiber. We used a similar method to perform measurements of the PRF: once the position of a nano-particle was found, we performed photon counting synchronized to the wave plate rotation, allowing the photon count (intensity) to be associated with each polarization state. We scanned 40 different QWP angles between 0\textdegree\  and  180\textdegree\  and 20 different HWP angles between 0\textdegree\ and  90\textdegree\ with equal increments. Finally, we verified nano-particle depositions after optical measurements had been completed by measuring the sample in a SEM and comparing measured particle positions with those detected via optical characterization.

Fig. \ref{fig:TheoryAndExp} shows a comparison of theoretical calculations and experimental results for a single GNS and a single GNR. Note that we performed the theoretical calculations using the parameters $\alpha$ (and $\theta_{\rm GNR}$ for nano-rods) measured by SEM after the optical measurements had been performed. In column (\textbf{a}), we show the SEM images of the particle on the optical nanofiber corresponding to the experimental results shown in columns (\textbf{c}) and (\textbf{d}). The 2D data shown in Figs. \ref{fig:TheoryAndExp}\textbf{(b)}, \textbf{(d)} are the theoretically predicted results and experimental results of the wave plate scan. The angles in Figs. \ref{fig:TheoryAndExp}(\textbf{c}) and (\textbf{e}) denote the location of the minimum value of the PRFs. The values of $\theta$ and $\phi$ for the experimental data are calculated by fitting the theoretical PRFs to the data using the least-square method. All data shown in Figs. \ref{fig:TheoryAndExp}\textbf{(c)}-\textbf{(d)} are normalized by the maximum value of the data set of $I_{+}$ or $I_{-}$. Comparison of theory and experimental data shows good qualitative and quantitative correspondence, indicating that our simple theoretical model is sufficient. In particular, our results for the single GNR case justify the assumption that the induced dipole moment of a single GNR is linearly polarized along the rod axis.

\begin{figure}[ht]
	\centering
	\includegraphics[width=\linewidth]{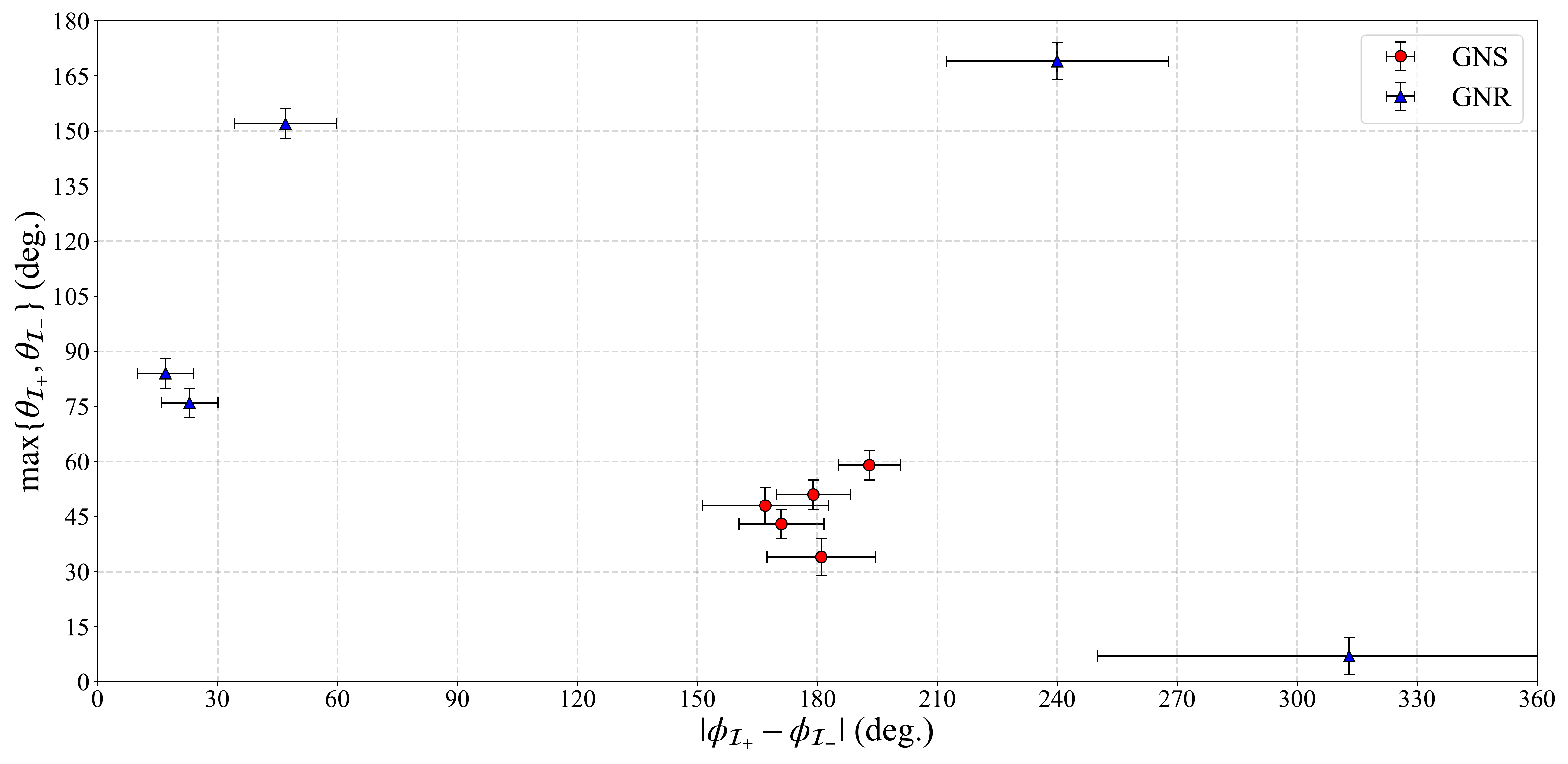}
	\caption{Comparison of parameters for single GNSs (red circle) and GNRs (blue triangles). Error bars show the estimated one standard deviation of the parameters extracted from the covariance of the fit to the data.}
	\label{fig:Sphere-Rod-Distribution}
\end{figure}

We now summarize the data measured for five single GNSs and five single GNRs on a single graph as shown in Fig. \ref{fig:Sphere-Rod-Distribution}. In particular, in each case we calculate the parameters $\mathrm{max}\{\theta_{\mathcal{I}_{+}},\, \theta_{\mathcal{I}_{-}}\}$ and $\lvert \phi_{\mathcal{I}_{+}} - \phi_{ \mathcal{I}_{-}} \rvert$ (where $\theta_{\mathcal{I}_{+}},\, \phi_{\mathcal{I}_{+}}$ and $\theta_{\mathcal{I}_{-}},\, \phi_{\mathcal{I}_{-}}$ are the positions on the Poincar\'{e} surface of the PRF minimum for $\mathcal{I}_{+}$ and $ \mathcal{I}_{-} $, respectively) for each data set and plot the former on the $y$ axis and the latter on the $x$ axis. The error bars show one standard deviation as calculated from the covariance of the fitted parameters using the least squares method. Considering these parameters for the cases of a single GNS, their values should lie along a vertical line at 180\textdegree\ and below a horizontal line at 90\textdegree\ because $\mathcal{I}$ for the case of a single GNS takes the minimum at $(\theta < 90^{\circ},\, \phi = +90^{\circ}\, \mathrm{or} \, -90^{\circ})$ in principle. As can be seen in Fig. \ref{fig:Sphere-Rod-Distribution}, the parameters for single GNSs concentrate around the vertical line at 180\textdegree\ and do not cross the horizontal line at 90\textdegree. On the other hand, as mentioned in the first paragraph of this section, the minimum value of the PRF for a single GNR lies on the $H-D-V-A$ great circle of the Poincar\'{e} sphere, \textit{i.e.}, $(0^{\circ}<\theta_{\mathcal{I}_{\pm}}<180^{\circ},\, \phi_{ \mathcal{I}_{\pm}} = 0^{\circ}\, \mathrm{or}\, 180^{\circ})$. For this reason, the parameter $\mathrm{max}\{\theta_{\mathcal{I}_{+}},\, \theta_{\mathcal{I}_{-}}\}$ can exceed the horizontal line at 90\textdegree\ and $\lvert \phi_{\mathcal{I}_{+}} - \phi_{ \mathcal{I}_{-}} \rvert$ is either 0\textdegree\ or 360\textdegree\ for a single GNR in theory. Therefore, these results allow us to meaningfully distinguish between scattering from nano-spheres and nano-rods on the nanofiber surface. More quantitatively, we note that the exact value of $\lvert \phi_{\mathcal{I}_{+}} - \phi_{ \mathcal{I}_{-}} \rvert$ in the case of nano-rods strays significantly from 0\textdegree\ or 360\textdegree\ in some cases. A number of reasons for this exist: The effect of scattering from the fiber surface leading to a change in the effective polarization seen by the rod, and the non-ideal nature of the wave plates used in the experiment are the most likely causes. Any of these will cause failure of our simple theoretical model to some degree. In addition, we note that when $\theta=0^o$ or 180$^o$, the value of $\phi$ becomes ill-defined (i.e. any choice of $\phi$ is consistent with the data in this case), leading to the large error bars seen when theta approaches either of these values. Nonetheless, we note that the main qualitative predictions of our model hold and Fig. \ref{fig:Sphere-Rod-Distribution} shows that the polarization response is a useful way to distinguish nano-spheres and nano-rods on the nanofiber surface in practice.

\section*{Conclusion}
We measured and compared the polarization response of single gold nano-spheres and gold nano-rods using the scattered light from the nanoparticles which coupled to the fundamental modes of an optical nanofiber. The experimental results show good correspondence with the theoretical calculations both qualitatively and quantitatively. Comparing the distribution of the angular position of the PRF minimum on the Poincar\'{e} sphere for rods and spheres shows that the data sets are clearly separated in this space. Using the techniques we have demonstrated here, it is possible to distinguish isotropic- and anisotropic-shaped particles with a size below the diffraction limit optically in most cases.
We believe that the method developed here will be particularly useful in the development of hybrid nano-waveguide - plasmonic nano-particle systems, as, for example, could be useful for the realization of high brightness fiber coupled single photon sources

\section*{Funding}
Japan Society for the Promotion of Science (JSPS) (JP19H04668).

\section*{Acknowledgments}
Mark Sadgrove acknowledges support from JSPS KAKENHI (Grant no. JP19H04668) in Scientific Research on Inovative Areas "Nanomaterial optical-manipulation". This research was partially performed using the facilities of the Fundamental Technology Center, Research Institute of Electrical Communication, Tohoku University.

\section*{Disclosures}
The authors declare no conflicts of interest.

\bibliography{PRFofGNR}

\end{document}